\documentclass[prd,nofootinbib,preprintnumbers,preprint]{revtex4}
\usepackage{indentfirst}

\usepackage{bm}
\usepackage{amsmath}
\usepackage{mathrsfs}
\usepackage{amsfonts}
\usepackage{amssymb}
\usepackage{dsfont}
\usepackage{euscript}
\usepackage[hyperfootnotes=false]{hyperref}
\usepackage[dvipsnames]{xcolor}

\newcounter{example}[section]
\newcounter{remark}[section]
\newcounter{theorem}[section]
\newcounter{proposition}[section]
\newcounter{lemma}[section]
\newcounter{corollary}[section]
\newcounter{definition}[section]
\DeclareMathAlphabet{\mathpzc}{OT1}{pzc}{m}{it}

\def\theremark{\arabic{section}.\arabic{remark}}
\def\thetheorem{\arabic{section}.\arabic{theorem}}

\def\thedefinition{\arabic{section}.\arabic{definition}}

\newcommand{\frc}[2]{\small{\raisebox{2pt}{$#1$}\big/\raisebox{-2pt}{$#2$}}}    
\renewcommand*{\email}[1]{\footnote{Electronic address: \href{mailto:#1}{\nolinkurl{#1}} }}

\begin{document}
\title{  Petrov types, separability and generalized photon surfaces\linebreak of supergravity black holes}
\author{Dmitri Gal'tsov\email{galtsov@phys.msu.ru}}
\author{Aleksandr Kulitskii\email{av.kulitskiy@yandex.ru}}
\affiliation{Faculty of Physics, Moscow State University, 119899, Moscow, Russia}

\begin{abstract}
The vacuum and electrovacuum Einstein equations for spacetimes with two commuting Killing vectors can be solved by indirect methods of integrable systems. But if, in addition, the spacetime admits an irreducible Killing tensor and the corresponding Klein-Gordon equation is separable, they can be integrated directly by separation of variables, as shown by Carter in 1968. We generalize this approach to supergravity and derive a metric ansatz that ensures the above properties for Petrov-type $I$. Our derivation is based on the Benenti-Francavilla ansatz for metrics admitting two commuting Killing vectors and an irreducible Killing tensor. We find additional constraints that guarantee the existence of two shear-free null geodesic congruences and the separability of the Klein-Gordon equation. The resulting class of metrics belongs to a certain sector of Petrov type $I$, called $I_B$, whose algebraically special subsector contains only type $D$. For this class, a direct integration of the supergravity equations seems possible. We also show that these spacetimes admit a general description of the generalized photon and massive particle surfaces recently introduced in connection with black hole shadows.
\end{abstract}
 \maketitle
\setcounter{page}{2}

\setcounter{equation}{0}
\setcounter{subsection}{0}
\setcounter{section}{0}
\setcounter{equation}{0}
\setcounter{subsection}{0}
\setcounter{section}{0}

\section{Introduction}\label{intro}
Hidden symmetries of spacetime, exhibited by second-rank Killing tensors, play no less an important role than isometries; for a review, see \cite{Cariglia:2014ysa,Frolov:2017kze,Lindstrom:2022nrm}. Over the past decade, significant progress has been made in the analysis of geometries admitting Killing-Yano (KY) tensors \cite{Frolov:2017kze,Lindstrom:2022nrm,Chervonyi:2015ima}, which are the strongest symmetries in the Killing hierarchy for $D$-type spacetimes. 

Hidden symmetries of spacetimes beyond type $D$, which do not allow for  KY structures, have also been examined, but their analysis is not yet so complete. Here we would like to investigate hidden symmetries of black holes in supergravities ${\cal N}=4, 8$, which include scalar fields and belong to general Petrov type $I$. The exact solutions for stationary black holes in these theories are fairly well known, largely due to hidden symmetries of a different kind that arise when these theories are dimensionally reduced in stationary spacetimes. As is well known, the dimensional reduction of supergravities to three dimensions leads to sigma models on homogeneous target  spaces \cite{Breitenlohner:1987dg,Breitenlohner:1998cv,Bouchareb:2007ax}.
The target space isometries include Harrison transformations, which allow generating  charged supergravity black holes from the Kerr metric. Apart from static solutions, which are easily obtained directly from the Einstein equations, most of the known stationary solutions were found just in this way, see e.g. \cite{Galtsov:1994pd,Bergshoeff:1996gg,Youm:1997hw,Bogush:2020obx}. A few exceptions are the supersymmetric solutions obtained using the Bogomolny equations \cite{Khuri:1995xk,Bellorin:2005zc} or solving the null geodesic equations in the target space \cite{Clement:1996nh} (also known as the nilpotent orbit method \cite{Bossard:2009we}).

Meanwhile, in vacuum and electrovacuum gravity, Carter showed that direct integration of stationary axisymmetric Einstein equations is possible for spacetimes admitting  both Hamilton-Jacobi and Klein-Gordon separability \cite{Carter:1968ks}. By explicit integration he found several classes of solutions, among which were Kerr and Kerr-Newman black holes of the Petrov type $D$ (for some details of the electromagnetic dressing see also \cite{Gair:2007uv}).  
 
Black holes in four-dimensional extended supergravities ${\cal N}= 4,\,8$ generically belong to  Petrov type $I$, though they still admit the second rank Killing tensor like their type $D$ cousins in vacuum and  electrovacuum gravity. We use the Benenti-Francaviglia (BF) \cite{Benenti:1979erw} parameterization of metrics admitting a Killing tensor independently of their Petrov type (Section 2).  We then require  for them the fulfillment of another important property of stationary axisymmetric spacetimes which holds in type $D$: the existence of two null geodesic shearfree congruences (Section 3). This is accomplished by imposing two additional conditions on the BF ansatz. 
Using the Newman-Penrose formalism, we find that the property of allowing null geodesic shearfree congruences extends beyond type $D$ also to a certain class of metrics of type $I$.
 
We then impose a third constraint, ensuring the separability of the Klein-Gordon equation (Section 4). The resulting class of metrics, which we call $I_B$, turns out to be similar to that used by Carter to perform direct integration of Einstein equations in vacuum and electrovacuum gravity  \cite{Carter:1968ks}. But our ansatz is now valid for type $I_B$ and is not constrained by the assumptions about the sources of matter used by Carter. This opens the way to a direct integration of the supergravity equations for this class.
Leaving an explicit integration  for a separate publication, here we verify that all explicitly known solutions for black holes do admit a simple polynomial description in terms of the constrained BF ansatz.
 
Section 5 is devoted to deriving conditions for the BF metrics to belong to type D. In Section 6 we show that for our class of metrics it is possible to have a unified description of the  photon  and massive particle surfaces, recently introduced \cite{Kobialko:2021aqg,Kobialko:2022ozq} as a tool for analyzing black hole shadows \cite{EventHorizonTelescope:2019dse,EventHorizonTelescope:2022wkp}. This follows from a special property of the BF-Killing tensor, which was called slice-reducibility. Our treatment is valid for all basic black holes in extended supergravities.
Finally, in Section 7 we present an explicit description of the known black hole solutions   in terms of BF functions and give explicit MPS equations for them.
 
\section{General Benenti-Francaviglia ansatz}
Our starting point is the class of four-dimensional metrics admitting a pair of commuting Killing vectors. More specifically, we will be interested in the stationary axisymmetric orthogonally transitive spacetimes (SAS) that can be parameterized by a block-diagonal metric, one of whose blocks is spanned by Killing vectors. For parametrization of the SAS spacetime that guarantees the separability of the Hamilton-Jacobi equation we use an ansatz given by Benenti and Francaviglia \cite{Benenti:1979erw}, based on Benenti's theorems \cite{Benenti:1976, Benenti1977SeparableDS} proving that in any dimension a necessary and sufficient condition for this to happen is the existence of a commuting system of Schouten-Nijenhuis brackets \cite{Schouten:1940,Nijenhuis:1955} (for a later discussion, see \cite{Cariglia:2011yt}) for Killing vectors and Killing tensors with a total number equal to the dimension of the spacetime. In this algebra, certain conditions on the eigenvectors of the Killing tensors must also be satisfied.In the four-dimensional case of SAS metrics, to which we restrict ourselves here, one (trivial) Killing tensor is the metric itself, so one irreducible Killing tensor is needed for separability.

Recently, SAS spacetimes with prescribed hidden symmetry have attracted attention in the search for viable alternatives to the Kerr metric for astrophysical modeling \cite{Papadopoulos:2018nvd,Papadopoulos:2020kxu,Carson:2020dez,Konoplya:2018arm}. In these applications, it was found, in particular, that  separation of variables in the Klein-Gordon equation for general BF metrics is not guaranteed.

A general BF parametrization consists of ten arbitrary functions, each depending on one variable. The  metrics are off-shell in the sense that Einstein's equations are not imposed. The
SAS  metric is written in the coordinates $x^\mu=(x^a,\,x^i)$, where $x^a=t,\, \varphi$ correspond to the subspace spanned by the Killing vectors $K^{(t)}=\partial_t$  and  $K^{(\varphi)}=\partial_\varphi$ and $x^i=r,\,y$, belong to orthogonal two-dimensional space whose metric without loss of generality can be assumed diagonal. BF ansatz looks somewhat simpler in terms of the contravariant metric tensor $g^{\mu\nu}=\left(g^{ab},\;g^{ij}\right)$ as follows: 
\begin{equation}\label{MetrUpBeneti}
g^{ab}=\Sigma^{-1}
\begin{pmatrix}
  A_3-B_3 \; & \; A_4-B_4\\   A_4-B_4\; & \;A_5-B_5
  \end{pmatrix}
,\qquad g^{ij}=-\Sigma^{-1}\begin{pmatrix}
  A_2&0\\  0& B_2
  \end{pmatrix}  
\end{equation}
where two sets of arbitrary functions are introduced $A_k(r),\,B_k(y),$ $k=1..5$ depending each on one variable, $r$ and $y$ respectively. In order to ensure existence of an exact Killing tensor (EKT), the conformal factor  $\Sigma=\Sigma(r,y)$ must be of the special form
\begin{equation}\label{Killing uslovie}
    \Sigma=A_1+B_1.
\end{equation}
Then the Killing tensor, satisfying the equation
\begin{equation}
\label{KillingTensorEquation}
\nabla_{(\alpha}K_{\mu\nu)}=0,
\end{equation} 
where symmetrization over indices is understood, also has a block diagonal form
\mbox{$K^{\mu\nu}=\left(K^{ab},\;K^{ij}\right)$ }, where
\begin{equation}\label{Killing}
\Sigma K^{ab}= \begin{pmatrix}
A_1B_3+A_3B_1\;&\;A_1B_4+A_4B_1  \\   A_1B_4+ A_4B_1\;&\;A_1B_5+A_5B_1
  \end{pmatrix},\qquad \Sigma K^{ij}= \begin{pmatrix}
  -A_2B_1&0\\  0& A_1B_2 \end{pmatrix}
\end{equation} 
The inverse metric (\ref{MetrUpBeneti}) and the Killing tensor (\ref{Killing}) have the following automorphism:
\begin{equation} \label{eq:automorphism}    A_1\leftrightarrow -B_1,
    \quad A_i\leftrightarrow B_i,\quad (i=2..4),
    \quad g^{rr}\leftrightarrow -g^{yy},\quad K^{rr}\leftrightarrow -K^{yy}.
\end{equation}
For an arbitrary conformal factor $\Sigma$ only a conformal Killing tensor may exist, we will come back to this later.
We keep separate notation for the conformal factor for further convenience.

Two blocks for the  covariant metric tensor $ 
g_{\mu\nu}=\left(g_{ab},\;g_{ij}\right)$: read  
\begin{equation}\label{MetrDownBeneti}
g_{ab}=\frac{\Sigma}{{\mathcal P}} \begin{pmatrix}
  A_5-B_5 \; & \; -A_4+B_4\\   -A_4+B_4\; & \;A_3-B_3
  \end{pmatrix},\qquad g_{ij}= -\Sigma  \begin{pmatrix}
  A_2^{-1}&0\\  0& B_2^{-1}
  \end{pmatrix}  
\end{equation}
  where 
\begin{equation}\label{detP}
   \mathcal P=(A_3-B_3)(A_5-B_5)-(A_4-B_4)^2. 
\end{equation}   
We shall assume that in a significant region of space-time (e.g. beyond the horizon or the ergosphere) all the BF coefficient functions are positive, and we shall not consider analytic continuation into the negative region, which of course can be done in the usual way. Therefore we shall often use square roots of the BF coefficients, assuming that they are real.  

Other sign conditions follow from the metric signature with the same reservations:
\begin{equation}
    A_3-B_3>0,\quad A_5-B_5<0,\quad {\cal P}<0.
\end{equation}
We also assume that the conformal factor $\Sigma$ is positive in the case where it is not assumed to be given by the formula (\ref{Killing uslovie}).
 
The static limit corresponds to 
\begin{equation}
    A_4\equiv 0,\quad B_4\equiv 0.
\end{equation}

One may wonder what is the gauge freedom inside the BF ansatz. The form of the metric suggests possibility of two coordinate transformations 
 \begin{equation}
     r\rightarrow \tilde{r}(r),\quad y\rightarrow \tilde{y}(y),
 \end{equation}
containing two arbitrary functions of independent variables. These can be used to impose additional conditions on $A_2,\; B_2$, e.g., $A_2=1=B_2$ (conformally flat metric in the  $r,y$ block). Other useful conditions, which turn out to be satisfied by all known supergravity black holes read as follows:
\begin{equation}\label{ab}
A_2  A_5 =  a^2={\rm  const},\quad   B_2  B_5 =  b^2={\rm  const}.
\end{equation}
In what follows we will assume the validity of this gauge.  
\subsection{Null tetrad}
We proceed by choosing some natural Newman-Penrose (NP) null tetrad \cite{Newman:1961qr} for the inverse metric tensor(\ref{MetrUpBeneti}):  \begin{align}
    g^{\mu\nu}=l^\mu n^\nu+n^\mu l^\nu- m^\mu \bar m^\nu-\bar m^\mu m^\nu.
\end{align}
We choose real vectors of the tetrad symmetrically:
\begin{subequations}
\label{bed tetrad}
\begin{align}
    &l={1}/{\sqrt{2\Sigma}}
    \left(\!\!\sqrt{\!A_3}\,\partial_t+C\,\partial_\varphi-\sqrt{\!A_2}\,\partial_r\right), &&
    m={1}/{\sqrt{2\Sigma}}\left(\!\!\sqrt{\!B_3}\,\partial_t+D\,\partial_\varphi-i\sqrt{\!B_2}\,\partial_y\right),\\
    &n={1}/{\sqrt{2\Sigma}}
    \left(\!\!\sqrt{\!A_3}\,\partial_t+C\,\partial_\varphi+\sqrt{\!A_2}\,\partial_r\right), &&
    \bar m={1}/{\sqrt{2\Sigma}}\left(\!\!\sqrt{\!B_3}\,\partial_t+D\,\partial_\varphi+i\sqrt{\!B_2}\,\partial_y\right).
\end{align}
\end{subequations}
where 
\begin{align}\label{CD}
    ({A_3\!-\!B_3})\,C\!=\! {\sqrt{\!A_3} \!\left(A_4\!-\!B_4\right)\!+\!\sqrt{\!B_3}\sqrt{-\mathcal P}},\qquad
   ({A_3\!-\!B_3})\,D\!=\!{\sqrt{\!B_3} \!\left(A_4\!-\!B_4\right)\!+\!\sqrt{\!A_3}\sqrt{-\mathcal P}}.
\end{align}
Recall that so far the number of arbitrary Benenti functions are ten.

\section{Null shear-free geodesic congruences}
Starting from the general BF class, we would like to select a subclass with additional properties noticed for some supergravity black holes \cite{Keeler:2012mq}. Vacuum solutions of type $D$ have the property, according to the Goldberg-Sachs theorem \cite{goldberg1962theorem}, of admitting two null geodesic congruences without shear. Electrovacuum
black holes arising in pure ${\cal N}=2$ supergravity without matter multiplets are also of type $D$ and also have similar congruences. Recall that stationary charged black holes in ${\cal N}=2$ supergravity can be obtained by Harrison transformations in the three-dimensional sigma model description \cite{Breitenlohner:1987dg} from the Kerr vacuum metric, so it is likely (although no actual proof has been given) that Harrison transformations preserve this property as well as geodesic integrability. Black hole solutions in extended supergravities with scalar moduli ${\cal N}=4,\,8$,  can also be obtained by Harrison transformations from Kerr metric \cite{Galtsov:1994pd,Cvetic:2017zde,Chow:2013tia,Chow:2014cca}, so we can expect them to share both of the above properties (as mentioned e.g. in \cite{Keeler:2012mq} for the two-charge STU solution). But Harrison transformations certainly do not preserve the Petrov type of the metric, since supergravity black holes with scalar fields generically belong to type $I$.

Thus, first of all we look for a reduced BF ansatz ensuring existence of two null geodesic shear-free congruences.
\subsection{First constraints }
Using algebraic computing and hints  from the known supergravity black hole solutions, one is led to consider the following constraints excluding two of ten BF functions:  
\begin{align}
\label{435}
A_4=\sqrt{A_3 A_5} ,\qquad B_4=\sqrt{B_3 B_5}.
\end{align}
With these conditions the functions entering  (\ref{CD}) simplify to
\begin{equation}
\sqrt{-\mathcal P}=\sqrt{A_3B_5}-\sqrt{A_5B_3},\quad \Rightarrow \quad
 C= \sqrt{A_5},\quad   D=\sqrt{B_5}.
\end{equation}
This reduced   BF ansatz leads to significant simplification of the null  tetrad:
\begin{align}
\nonumber
    &l={1}/{\sqrt{2\Sigma}}    \left(\!\!\sqrt{\!A_3}\,\partial_t+\sqrt{\!A_5}\,\partial_\varphi-\sqrt{\!A_2}\,\partial_r\right), &&
    m={1}/{\sqrt{2\Sigma}}\left(\!\!\sqrt{\!B_3}\,\partial_t+\sqrt{\!B_5}\,\partial_\varphi-i\sqrt{\!B_2}\,\partial_y\right),\\
    \label{BT}
    &n={1}/{\sqrt{2\Sigma}}    \left(\!\!\sqrt{\!A_3}\,\partial_t+\sqrt{\!A_5}\,\partial_\varphi+\sqrt{\!A_2}\,\partial_r\right), &&
    \bar m={1}/{\sqrt{2\Sigma}}\left(\!\!\sqrt{\!B_3}\,\partial_t+\sqrt{\!B_5}\,\partial_\varphi+i\sqrt{\!B_2}\,\partial_y\right).
\end{align}

In the static limit our relations (\ref{435}) degenerate, so we have to be careful in making correct assignments for the remaining BF coefficients. Looking at the constrained tetrad \ref{BT}, we realize that a correct choice in the static limit will be
\begin{equation}
    A_5\equiv 0,\quad B_3\equiv 0,\quad A_3\neq 0,\quad B_5\neq 0,
\end{equation}
implying $a=0, \,b\neq 0.$

\subsection{Newman-Penrose analysis} To understand which restrictions on the nature of spacetime is put by the first constraints, 
let's continue our NP analysis. 
Recall the definitions of the NP projections of the covariant derivatives
\begin{equation}\label{NPder}
D=l^\mu\nabla_\mu,\quad \Delta=n^\mu\nabla_\mu,\quad \delta=m^\mu\nabla_\mu,\quad {\bar \delta}=\bar m^\mu\nabla_\mu,
\end{equation}
and the action of $D, \Delta$ on the vectors $l^\mu,\, n^\mu$
\begin{align}
    &Dl^\mu= (\epsilon+\bar\epsilon)l^\mu-{\bar\kappa} m^\mu-\kappa{\bar m}^\mu   \label{l}     \\
    &\Delta n^\mu=-(\gamma+\bar\gamma)n^\mu+  \nu m^\mu +{\bar\nu}{\bar m}^\mu  \label{n}. 
\end{align}
Consider null congruences aligned with $l^\mu,\,n^\mu$. If $\kappa=0=\nu$  they are {\em geodesic}, with $\epsilon ,\;\gamma$ being measure of non-affinity. Another important quantity of null congruences is {\em shear}, which is defined for them as 
\begin{equation}
  \sigma=-m^\mu\delta \,l_\mu,\qquad\qquad  \bar\lambda=m^\mu\delta\, n_\mu 
\end{equation}
respectively.
Calculating the spin coefficients (see Appendix A) for our tetrad, we find:
\begin{equation} \label{sc}
\kappa=\nu=0,\qquad\qquad\sigma=\lambda=0,
\end{equation}
which means that both congruences are {\em geodesic} and {\em shearfree}.
Other spin coefficients are generically non-zero and pairwise equal:
\begin{equation}
\label{spin}
\mu=\rho,\quad\tau=\pi,\quad\epsilon=\gamma,\quad\alpha=\beta. 
\end{equation}
 
Such properties are typical for Petrov type $D$, once two congruences are principal null directions of the Weyl tensor (which is also the case, as we shall see shortly). To establish the Petrov type in our case, we calculate the NP projections of the Weyl tensor:
 \begin{align}
&\Psi_0=-C_{\alpha\beta\gamma\delta} l^\alpha m^\beta l^\gamma m^\delta,  \nonumber\\
&\Psi_1=-C_{\alpha\beta\gamma\delta} l^\alpha n^\beta l^\gamma m^\delta,  \nonumber\\
&\Psi_2=-C_{\alpha\beta\gamma\delta} \left(l^\alpha n^\beta l^\gamma n^\delta -l^\alpha n^\beta m^\gamma  \bar{m}^\delta\right)/2, \\
&\Psi_3=-C_{\alpha\beta\gamma\delta} n^\alpha l^\beta n^\gamma m^\delta,  \nonumber\\
&\Psi_4=-C_{\alpha\beta\gamma\delta} n^\alpha \bar{m}^\beta n^\gamma \bar{m}^\delta,   \nonumber
\end{align}
From the computer assisted calculations  one finds that two of them are zero, 
\begin{equation}
 \Psi_0=0=\Psi_4,   
\end{equation} while the others are rather cumbersome in terms of BF coefficients.  Still one can extract the following  relation between the other two:
\begin{equation}
\label{psi} \Psi_1=\Psi_3, 
\end{equation}
reflecting obvious symmetry of the tetrad with respect to interchange $A\leftrightarrow B$.
Vanishing of $\Psi_0$ and $\Psi_4$ means that the real vectors $l^\mu,\;n^\mu$
are two distinct principal null directions of the Weyl tensor for a constrained BF metric. At the same time, this means that our tetrad is not canonical for determination of the Petrov type. 
We therefore proceed by computing
the values of the quadratic and cubic curvature invariants of the Weyl tensor
\begin{equation}
 I=\Psi_0\Psi_4-4\Psi_1\Psi_3+3\Psi_2^2,\quad\quad J=  \begin{vmatrix}
\Psi_4 &\Psi_3 & \Psi_2\\
\Psi_3 &\Psi_2 & \Psi_1\\
\Psi_2 &\Psi_1 & \Psi_0
\end{vmatrix}.
\end{equation}
As is known, in order for a metric to be algebraically special, the following relationship between these invariants must be satisfied:
\begin{equation}
    I^3=27 J^2.
\end{equation}
It is easy to see that (\ref{psi}) implies that  the constrained BF metrics are algebraically special if in addition to (\ref{psi})  the following conditions hold 
\begin{equation}\label{pet}
  \Psi_1^2=k \Psi_2^2,  \quad {\rm with}\quad  k=9/16,\;\; {\rm or} \;\; 0.  
\end{equation}
If (\ref{pet}) does not hold, the metric is of the Petrov type I. If  (\ref{pet}) if satisfied with $k=0$ and $\Psi_2\neq 0$ then the metric type is $D$. Other algebraically special types are not possible within the constraint (\ref{psi}), for example if one supposes type II, for which $\Psi_0=0=\Psi_1$, one immediately finds that the Weyl tensor completely vanishes, i.e. the   metric is of type $O$.  

By Goldberg-Sachs theorem, the {\em vacuum} spacetime is algebraically special if it contains a null geodesic shear-free congruence. If there are two such congruences, the Petrov type is $D$. Therefore in the case of type $I$ spacetime, admitting a null geodesic shear-free congruence, the Ricci tensor should be non-zero. This is the case for supergravity black holes. Thus our class   $I_B$ consists of non-vacuum metrics, possessing a Killing tensor and a pair of null geodesic shear-free congruences. These properties are close to properties of $D$ type, they will ensure separability of Hamilton-Jacobi equation and Klein-Gordon equation if some further conditions on the Ricci tensor are fulfilled (see below). However a Killing-Yano tensor exists only in the algebraically special case $D$, so the Dirac equation generally is non-separable.

Note that in the null tetrad (\ref{BT}) the vectors $l^\mu,\,n^\mu$ are not affinely parameterized (the spin coefficients $\epsilon,\,\gamma$ being non-zero). One can pass to an affinely parameterized congruence $l^\mu$ (like in the case of the Kinnersley tetrad for Kerr metric)
by rescaling the tetrad vectors as
\begin{equation}
  l^\mu \rightarrow f^{-1}(r,y) \;   l^\mu,\quad   n^\mu \rightarrow f(r,y) \;  n^\mu,\quad f=\sqrt{{\Sigma}/{A_2}}.
\end{equation}
This leaves unchanged the coefficients (\ref{sc}) and the Weyl projections (\ref{psi}), so our conclusions will be the same.

One can also use the classification scheme based on expansion of the Weyl tensor in terms of  bivectors. This leads to the traceless symmetric $3\times 3$ matrix $\bf{Q}$, related to Weyl projections via
\begin{equation}
\bf{Q}=  \begin{pmatrix}
\Psi_2-\frac12(\Psi_0+\Psi_4) &\frac12i(\Psi_4-\Psi_0) & \Psi_1-\Psi_3\\
\frac12i(\Psi_4-\Psi_0) &\Psi_2+\frac12(\Psi_0+\Psi_4) & i(\Psi_1+\Psi_3)\\
 \Psi_1-\Psi_3 &i(\Psi_1+\Psi_2) & -2\Psi_2
\end{pmatrix}.
\end{equation}
With our conditions (\ref{psi}) this matrix becomes
\begin{equation}
\bf{Q}=  \begin{pmatrix}
\Psi_2 &0 &0\\
0&\Psi_2  & 2i\Psi_1 \\
 0 &2i\Psi_1  & -2\Psi_2
\end{pmatrix}.
\end{equation}
The case of three different eigenvalues corresponds to type $I$, one degenerate
eigenvalue (in the case $\Psi_1=0$) lead to type $D$.

Thus, we have proved that the   BF metrics (\ref{MetrUpBeneti}) with the additional conditions (\ref{435}) defines a special Petrov type $I_B$ class, which shares with the type $D$ two important properties: 1) it admits two independent null shearfree geodesic congruences, 2) it has an irreducible Killing tensor of the second rank.
In terms of the constrained tetrad (\ref{BT}) the Killing tensor has only two non-vanishing Newman-Penrose 
 projections exactly as in the case of the type $D$ metrics:
\begin{align}
\label{KC} K_{ln}=B_1, \qquad\qquad \qquad\quad K_{m\bar m}=A_1.
\end{align}
This gives some geometric interpretation of two Benenti coefficients entering the conformal factor 
(\ref{Killing uslovie}). 

We further show that the known solutions for black holes in ${\cal N}=2,\,4,\,8$ extended supergravity theories belong to this class $I_B$, which ensures the separability of the Hamilton-Jacobi equation. 
Moreover, they satisfy the third constraint, which we are going to establish, ensuring separability of the Klein-Gordon equation.

\subsection{ Separability of geodesic equations}

Separability of the Hamilton-Jacobi equation 
\begin{equation}
  \frac{\partial S}{\partial x^\mu}  \frac{\partial S}{\partial x^\nu}\; g^{\mu\nu}=\mu^2 
\end{equation}
can be easily demonstrated. Assuming \begin{equation}\label{SEL}
   S=-Et+L\varphi+S_r(r)+S_y(y), 
\end{equation}
denoting $p_\mu=\partial S/\partial x^\mu$, and taking into account that all 
$A_k$ depend  only on $r$, while all $B_k$ depend  only on $y$,  we obtain
\begin{align}\label{HJsepar}
 &A_2\,p_r^2+U_r=0,\qquad  \qquad\qquad B_2\,p_y^2+U_y=0,\\ \label{Ury}
 &U_r= {\cal C}- \varepsilon_a^2 +A_1 \mu^2  ,\qquad \qquad U_y=  -{\cal C}+\varepsilon_b^2 +B_1 \mu^2 , \\ \label{epsiab}     
&\varepsilon_a=E\sqrt{A_3}-L\sqrt{A_5},\qquad \quad \,\varepsilon_b=E\sqrt{B_3}-L\sqrt{B_5},  
\end{align}
where ${\cal C}$ is the Carter separation constant related to the Killing tensor as
\begin{equation}
   {\cal C}=p_\mu p_\nu  \,K^{\mu\nu}=\Sigma^{-1}\left[ A_1\left( \varepsilon_b^2 +B_2 p_y^2 \right)+B_1\left( \varepsilon_a^2 -A_2 p_r^2 \right)\right].
\end{equation}
Note that the above definition of potentials is such that Carter's constant enters additively. This has certain advantages when discussing spherical and conical orbits.
The radii of spherical orbits, e.g., correspond to solutions of the equations
\begin{equation}\label{UUder}    U_r=0=U'_r.
\end{equation}
The derivative equation does not contain the Carter integral, and thus define radii of spherical orbits as functions of $E,\,L$. Then substitution of this radius into the equation $U=0$ will show, that the Carter integral for spherical orbits is a function of Killing vector integrals.
 
\subsection{Variable mass}
In conclusion to this section we explore in which case separability of the Hamilton-Jacobi equation can be extended to the case of variable coordinate-depending mass.  This may be of  interest in the study of shadows of black holes surrounded {\em by plasma} \cite{Tsupko:2013cqa,Badia:2021kpk,    Bezdekova:2022gib, Perlick:2023znh}. A photon in electron plasma acquires an effective mass determined by the refractive index, which depends on the coordinates through the electron density. Since the latter in the vicinity of black holes depends on the coordinates,  the propagation of photons in the geometrical optics approximation is described by the Hamilton-Jacobi equations with mass depending on the coordinates:
$$\mu^2\Rightarrow {\cal M}^2(x).$$
Separation of variables in the Hamilton-Jacobi equation with variable mass is possible only if
\begin{equation}
  {\cal M}^2(x)=\frac{{\cal M}_r^2(r)+{\cal M}_y^2(y)}{A_1+B_1}.  
\end{equation}
This is consistent with what was found for the Kerr metric in \cite{Perlick:2023znh}.
\section{Separability of Klein-Gordon equation} Generic type $I_B$ class of metrics still does not quarantee separability of the wave equations. Consider the Klein-Gordon equation for a real scalar field $\phi$:
\begin{equation}
\label{KG}
\Box \phi= \frac1{\sqrt{-g}}\partial_\mu\left(\sqrt{-g}g^{\mu\nu}\partial_\nu \phi\right)=-\mu^2\phi. 
\end{equation}
The metric determinant is crucial for separability. It is easier to calculate it in BF form using contravariant components (\ref{MetrUpBeneti}). One obtains
\begin{equation}\label{third}
  \sqrt{-g}=\frac{\Sigma^2}{\sqrt{A_2B_2}\sqrt{-\cal P}}  
\end{equation}
With the first two constraints (\ref{435}), one can use (\ref{detP}) to obtain $\cal P$, leading to
\begin{equation}\label{det}
  \sqrt{-g}=\frac{\Sigma^{2}}{\sqrt{A_2 B_2}\left(\sqrt{A_3B_5}-\sqrt{A_5B_3 }\right)}.  
\end{equation}
Considering the inverse metric (\ref{MetrUpBeneti}), it becomes clear that the condition for separability is
$$\sqrt{-g}=\Sigma.$$
This leads to the third constraint on the Benenti coefficient functions
\begin{equation}\label{sigCons}   \Sigma= \sqrt{A_2 B_2}\left(\sqrt{A_3B_5}-\sqrt{B_3 A_5}\right),
\end{equation}
which can also be rewritten as 
\begin{equation}\label{sigConsAlt}    A_1+B_1= bA_{23}-aB_{23},
\end{equation}
where we introduced \begin{equation}
A_{23}=\sqrt{A_2A_3},\quad\quad B_{23}=\sqrt{B_2B_3},
\end{equation} and used the gauge (\ref{ab}).  Differentiating this with respect to the relevant arguments, one can find useful differential relations following from the third constraint:
\begin{equation}
   \label{A1B1} A_1'=bA_{23}',\qquad B_1'=aB_{23}'.
\end{equation}
(Note that primes cannot be omitted in these ratios!)
With the help of (\ref{third}) it is easy to establish the separability of equation (\ref{KG}). Substituting the product
\begin{equation}
  \phi(x^\mu)={\rm e}^{-i\omega t +i m\varphi}  R(r) Y(y),
\end{equation}
and dividing  the Eq.(\ref{KG})  by $\phi$ we obtain:
\begin{equation}\label{KGsep}
\frac{((A_2)'R)'}{R} +   \frac{((B_2)'Y)'}{Y} +U(r)-V(y)=0,  
\end{equation}
where the primes denote derivatives with respect to the corresponding arguments $r,\,y$, and the potential terms are equal to 
\begin{align}
    &U(r)=(\omega\sqrt{A_3}-m\sqrt{A_5})^2- \mu^2 A_1,\\ &V(y)=(\omega\sqrt{B_3}-m\sqrt{B_5})^2+   \mu^2 B_1.
\end{align}
The separability of Eq. (\ref{KGsep}) is obvious.
The third constraint effectively reduces the number of arbitrary functions to seven, two of which $A_2, B_2$ can still be fixed using the gauge freedom, so the number of essentially independent functions is five.

The separability of the Klein-Gordon equation can also be investigated using the second-order Carter differential operator associated with the Killing tensor: 
\begin{equation}
  \hat{K}=\nabla_\mu K^{\mu\nu}\nabla_\nu,
\end{equation}
which must commute with D'Alembert operator \cite{Keeler:2012mq}. 
The commutator was elaborated in \cite{Carter:1977pq}:
\begin{align}
\label{KommutatorKilling}
    [\square,\hat K]\,\phi=\frac{4}{3}\nabla_\alpha(K_\sigma{}^{[\alpha}R^{\beta]\sigma})\nabla_\beta\,\phi.
\end{align}
Projecting the tensor on the right side onto the NP tetrad, we obtain
\begin{align}
\nonumber  K_\sigma{}^{[\alpha}R^{\beta]\sigma}&=2(K_{ln}+K_{m\bar m})
    (n^\beta(\bar m^\alpha \Phi_{01}+m^\alpha\Phi_{10})-n^\alpha(\bar m^\beta \Phi_{01}+m^\beta\Phi_{10})+\\
    &+(l^\beta\bar m^\alpha-l^\alpha\bar m^\beta)\Phi_{12}+(l^\beta m^\alpha-l^\alpha m^\beta)\Phi_{21}).
\end{align}
Thus, a sufficient condition for commutativity is that two Ricci scalars vanish:
\begin{equation}\label{deviators}
  \Phi_{01}=R_{\mu\nu}l^\mu m^\nu/2=\overline{{\Phi}_{10}} ,\qquad\qquad\Phi_{12}=R_{\mu\nu}n^\mu m^\nu/2=\overline{{ \Phi}_{21}}.  
\end{equation} 
Their expression, taking into account the third constraint (\ref{sigCons}) in terms of Benenti quantities, is given by formula (\ref{Phi01}) in Appendix B. Equating them to zero, we obtain:  
\begin{align}
\label{REQ}    a A_{23}''-b B_{23}''=0, 
\end{align}
where the primes denote derivatives with respect to the corresponding arguments. Given (\ref{A1B1}), this can also be rewritten as
\begin{equation}
    a^2A_1''-b^2B_1''=0.
\end{equation}
Since one term is a function of $r$ and the other is a function of $y$, each must be a constant. In other words, $A_1$ and $B_1$ must be at most quadratic polynomials of the corresponding arguments.

Let us give an NP-description of the operators $\Box$ and $\hat K$ in terms of derivatives
 (\ref{NPder})
and the spin coefficients:
\begin{align}
	\square&=(D+\varepsilon+\bar\varepsilon-\rho-\bar\rho)\Delta+(\Delta+\mu+\bar\mu-\gamma-\bar\gamma)D-\nonumber\\
\label{gordon}
	&-(\delta-\tau-\bar\alpha+\beta+\bar\pi)\bar\delta-(\bar\delta-\alpha-\bar\tau+\pi+\bar\beta)\delta,\\\hat K&=(D+\varepsilon+\bar\varepsilon-\rho-\bar\rho)K_{ln}\Delta+(\Delta+\mu+\bar\mu-\gamma-\bar\gamma)K_{ln}D+\nonumber\\
	&+(\delta-\tau-\bar\alpha+\beta+\bar\pi)K_{m\bar m}\bar\delta+(\bar\delta-\alpha-\bar\tau+\pi+\bar\beta)K_{m\bar m}\delta,
	\end{align}	
	It is useful to introduce the following set of operators:
\begin{subequations}	\label{stairs_operators}
	\begin{align}
	    &{} D_n^\pm = \sqrt{{A_3}/{A_2}} \,{\partial_t}\pm{\partial_r}+ \sqrt{A_5/A_2}\,
  \partial_\varphi \pm n\,{\partial_r(\ln A_2)},\\
     &{} L_s^\pm = \sqrt{B_3/B_2} \,{\partial_t}\pm i\,{\partial_y}+ \sqrt{B_5/B_2}\,
  \partial_\varphi\pm is\,{\partial_y(\ln B_2)},
	\end{align}
	\end{subequations}
	where $n$ and $s$ are integers. 
	These operators satisfy two simple identities:
	\begin{align}	\label{stairs_commutator}	    {} D_n^\pm A_2{}^k = A_2{}^k\,{} D_{n+ k }^\pm,\qquad {} L_s^\pm B_2{}^l = B_2{}^l\,{} L_{s+ l}^\pm.
	\end{align}
	In terms of these operators, the directional derivatives acting on a scalar function will have the form:
	\begin{align}	\label{stairs_different}
    D= \sqrt{A_2/ 2\Sigma}\,{} D_0^-,&& \Delta=\sqrt{A_2/ 2\Sigma}\,{} D_0^+,&&\delta=\sqrt{B_2/ 2\Sigma}\,{} L_0^-,&&    \bar\delta=\sqrt{B_2/ 2\Sigma}\,{} L_0^+.
	\end{align}
Also, taking into account the explicit form of the spin coefficients (\ref{AddSpinCoeff1}-\ref{AddSpinCoeff4}), we can obtain the following representations for the combinations included in the d'Alembertian:
	Also, taking into account the explicit form of the spin coefficients (\ref{AddSpinCoeff1}-\ref{AddSpinCoeff4}), we can obtain the following representations for the combinations included in the d'Alembertian:
 \begin{subequations}
     \begin{align}
         &\varepsilon+\bar\varepsilon-\rho-\bar\rho\;\,=\;\;\;\,\sqrt{A_2/ 8\Sigma^3}\left(2\Sigma
         \,\partial_r\ln\left[
         \sqrt{A_3B_5}-\sqrt{A_5B_3}
         \right]-3\,\partial_r\,\Sigma\right),\\
       &\bar\pi-\tau+\beta-\bar\alpha = -i\sqrt{B_2/ 8\Sigma^3}\left(2\Sigma
         \,\partial_y\ln\left[
         \sqrt{A_3B_5}-\sqrt{A_5B_3}
         \right]-3\,\partial_y\,\Sigma\right).
     \end{align}
 \end{subequations}
It is easy to see that in the general case, after substitution into the Klein-Gordon equation, the latter does not allow separation of variables unless the second constraint (\ref{sigCons}) is imposed. In this case, our expressions are simplified to
	\begin{subequations}
\label{gordon_coefficients}
	\begin{align}
   &\varepsilon+\bar\varepsilon-\rho-\bar\rho\;\, = \,\;\sqrt{A_2}\,\partial_r\,(2\Sigma)^{-1/2}-(2\Sigma)^{-1/2}\,\partial_r\sqrt{A_2},\\
	    &\bar\pi-\tau+\beta-\bar\alpha =i \sqrt{B_2}\,\partial_y\,(2\Sigma)^{-1/2}-i(2\Sigma)^{-1/2}\,\partial_r\sqrt{B_2}.
	\end{align}
	\end{subequations}	
Using Eqs.  (\ref{stairs_operators}-\ref{gordon_coefficients}), the d'Alembert operator (\ref{gordon}) can be cast into the form	\begin{align}\label{DalamUpDown}
	    \square=\frac{1}{2\Sigma}\left[A_2\left({} D_1^-{} D_0^++{} D_1^+{} D_0^-\right)-B_2\left({} L_1^-{} L_0^++{} L_1^+{} L_0^-\right)\right],
	\end{align}
	while the Carter operator can be presented as
	\begin{align}
	\hat K\! = \frac{1}{2\Sigma}\!\left[A_2B_1\!\left({} D_1^-{} D_0^+\!+\!{} D_1^+{} D_0^-\right)\!+\!B_2A_1\!\left({} L_1^-{} L_0^+\!+\!{} L_1^+{} L_0^-\right)\right]\!=\frac{B_2}{2}\!\left({} L_1^-{} L_0^+\!+\!{} L_1^+{} L_0^-\right)\!+\!B_1\square.
	\end{align}
By direct substitution one can verify that it commutes with the d'Alembertian (\ref{DalamUpDown}), taking into account the fact that the second term in the solution space $\square \,\phi=0$ vanishes.
 \subsection{ Separable supergravity backgroounds}
By imposing three constraints on the BF coefficients, we arrive at the following metric paremeterization:  
\begin{equation}
ds^2=\frac{A_2B_2}{\Sigma}\left( \sqrt{B_5}dt-\sqrt{B_3}d\varphi\right)^2  -  \frac{A_2B_2}{\Sigma}\left( \sqrt{A_5}dt-\sqrt{A_3}d\varphi\right)^2  -\frac{\Sigma}{A_2}dr^2-\frac{\Sigma}{B_2}dy^2,
\end{equation}
where
\begin{equation}
   \Sigma=\sqrt{A_2B_2}(\sqrt{A_3B_5}-\sqrt{A_5B_3}). 
\end{equation}
This is precisely the Carter metric ansatz \cite{Carter:1968ks}  for which Einstein's vacuum and electrovacuum equations were solved directly.
Several families of solutions were obtained among which  Kerr and Kerr-Newman black holes belonging to the Petrov type $D$ (for further applications see also \cite{Gair:2007uv})

Consider now the generic 4D supergravity bosonic action which is a special scalar-vector-tensor theory with multiple scalar and vector fields
\cite{Breitenlohner:1987dg,Breitenlohner:1998cv}. It includes   $n_s$ scalar moduli $\Psi^A$, $A=1,\ldots,n_s$ (dilatons and axions) and $n_v$ abelian vector fields $F^I=dA^I$, $I=1,\ldots,n_v$:
\begin{equation} \label{eq:general_action}
 S =
 \int d^4x\left[
    \left(
          R 
        - \frac{1}{2} f_{AB}\, \partial_\mu\Psi^A\partial^\mu\Psi^B
        - \frac{1}{2} K_{IJ} F^I_{\mu\nu} F^{J\mu\nu}
    \right) \sqrt{-g}
    - \frac{1}{2} H_{IJ} F^I_{\mu\nu} F^J_{\lambda\tau} \epsilon^{\mu\nu\lambda\tau}
  \right].
\end{equation} The  scalar moduli parametrize a four-dimensional coset (e.g. $U(8)/E_{7(7)}$ for  ${\cal N}=8$ supergravity) with an associated target metric $f_{AB}$. Vector fields transform under the same global symmetry implemented by real symmetric matrices $K_{IJ}$, $H_{IJ}$  depending on scalar fields $\Psi^A$ (summation over the repeated indices $I,\,J$ is understood).  
The corresponding  Einstein equations read:
\begin{equation}
\label{Ric}
    R_{\mu\nu} = 
      \frac{1}{2} f_{AB} \Psi_{,\mu}^A\Psi_{,\nu}^B 
    - K_{IJ} \left(
          F^I_{\mu\lambda}{F^{J\lambda}}_\nu
        + \frac{1}{4} g_{\mu\nu} F^I_{\alpha\beta}F^{J\alpha\beta}
    \right).
\end{equation}
The scalar fields depend only on $r,y$, so they contribute directly only in transverse part of the Ricci tensor. The Maxwell sector at the right hand side of this equations has the same structure as in pure Einstein-Maxwell system. All this ensure (the details will be given elsewhere) that supergravity Einstein equations will separate for our metric ansatz similarly to Einstein-Maxwell case \cite{Carter:1968ks}. Note, that the Smarr mass formulas for black hole solutions of the theory (\ref{eq:general_action}) was shown recently to repeat the case of the Einstein--Maxwell theory \cite{Bogush:2024lrd}.  

To our knowledge, no direct integration of the Einstein supergravity equations by separation of variables has been performed so far, although the leading solutions for black holes have been obtained using indirect methods. In addition to Harrison transformations, one can mention the guesswork of obtaining the BPS solution \cite{Bergshoeff:1996gg}
or the integration of the null geodesic equations for the target space of the sigma model \cite{Clement:1996nh} (also known as the nilpotent orbit method \cite{Bossard:2009we}).
The situation is  different in the electrovacuum case, where large classes of solutions were obtained by direct integration of the Ernst equations. 
 \section{ Killing-Yano and type D}
Here we find conditions on the BF functions that guarantee that the solution is algebraically special, which for our class $I_B$ means type $D$. In this case, the Killing-Yano tensor that guarantees the separability of the Dirac equation also exists. We will use this as a tool to find conditions for type $D$.
 
\subsection{Killing-Yano  }
The Killing-Yano tensor $Y_{\mu\nu}=-Y_{\nu\mu}$  satisfying
the equation
\begin{equation}\label{KYEquationInitial} 
\nabla_{(\alpha} Y_{\mu)\nu}=0, 
\end{equation}    can be regarded as a ``square root'' of the Killing tensor:
\begin{equation}\label{KJtoKill}
   Y_\mu{}^\alpha Y_\alpha{}_\nu=K_{\mu\nu}. 
\end{equation}
Since we know the Killing tensor (\ref{Killing}) independently of Petrov type of the metric, we may consider  Eqs.(\ref{KYEquationInitial}) and (\ref{KJtoKill}) as independent conditions which prescribe the metric to be of type $D$, and define the KY tensor itself. In NP description, our Killing tensor has only two non-zero components (\ref{KC}): 
$K_{ln}$ and $ K_{m\bar m}$. So projecting (\ref{KJtoKill}) on the NP tetrad, one obtains:
\begin{equation}
\label{KYC}
Y_{ln}^2=K_{ln},\qquad \qquad Y_{m\bar{m}}^2=- K_{m\bar m},
\end{equation}
the other NP components of the KJ tensor being zero. Together with Eq. (\ref{KC}) this gives
\begin{equation}
  Y_{ln}= p \sqrt{B_1},\qquad Y_{m\bar m}=i q \sqrt{A_1},\qquad     p=\pm 1,\qquad   q=\pm 1,   
\end{equation}
where we introduced sign factors to be fixed later.
\subsection{Consistency conditions for type D}
Now we have to satisfy the KY equation (\ref{KYEquationInitial}) for consistency, in other words, we have to find new constraint equations on the BF coefficient functions which ensure that solution belongs to type $D$. Projecting KJ equation (\ref{KYEquationInitial}) onto the NP tetrad we get sixteen equations, from which, with account for   (\ref{KC}) and   pairwise equality of the spin coefficients (\ref{spin}), only the following six are relevant:
 \begin{subequations}
 \begin{align}
 \label{KYS1}
    &(\tau+\bar\pi)Y_{ln}-(\tau-\bar\pi)Y_{m\bar m}=0,
    &&(\rho+\bar\rho)Y_{ln}-(\rho-\bar\rho)Y_{m\bar m}=0,\\
\label{KYS5}
    &DY_{m\bar m}-\rho (Y_{ln}-Y_{m\bar m})=0,
    &&\delta Y_{ln}-\bar \pi(Y_{ln}+ Y_{m\bar m})=0,\\
      \label{KYS6}
    &\Delta Y_{m\bar m}+\mu( Y_{ln}-Y_{m\bar m})=0,
    &&\delta Y_{ln}+\tau(Y_{ln}- Y_{m\bar m})=0.
    \end{align}
    \end{subequations}
For the spin coefficients involved, from (\ref{AddSpinCoeff1},\ref{AddSpinCoeff2}) with account for the constraint  (\ref{sigCons}) one obtains: 
    \begin{align}
        \mu=\rho=\frac{b}{2\Sigma}\sqrt{{A_2}/{2\Sigma}}\left( A_{23}'-i\, B_{23}'\right),&&\tau=\pi=\frac{ a}{2\Sigma}\sqrt{B_2/2\Sigma}\left(A_{23}'-i\,B_{23}'\right),
    \end{align}
Then from the first pair of equations of the system we get the following relation:
\begin{align}
     p \,A_{23}'\sqrt{B_1}= q  \,B_{23}'\sqrt{A_1},
\end{align}
which separates into pair of one-variable equations:  
\begin{align}
    pA_{23}'=2\sqrt{A_1}, &&qB_{23}'=2\sqrt{B_1}
\end{align}
where we introduced the coefficient two for further convenience. Now we have to satisfy the second and third pairs of equations (\ref{KYS5}, \ref{KYS6}). Taking into account definition of NP derivatives (\ref{NPder}), we obtain
\begin{align}
    (\sqrt{A_1})'={b}/{p},&&(\sqrt{B_1})'=-{a}/{q}
\end{align}
From (\ref{Killing uslovie}) and (\ref{sigCons}) follows that $$A_1+B_1=b A_{23}-a B_{23}.$$ To satisfy this,  one has to choose $p=1$ and $q=-1$,  Thus, we have found a system of restrictions on the BF coefficients that guarantee the existence of the KY tensors, i.e. the belonging of the metric to the $D$ type:
\begin{subequations}
    \begin{align}
    \label{KY2Cond}
   & {A_1'}=2 b\sqrt{A_1},&& {B_1'}=2a\sqrt{B_1};\\
   \label{KY3Cond}
   & A_{23}' ={2\sqrt{A_1}},&& B_{23}' =-{2\sqrt{B_1}}.
\end{align}
\end{subequations}
Intergration of this system provides generic form for some of the metric coefficients for type 
$D$ BF sector:
\begin{subequations}
\begin{align}\label{solForDa}
    &A_1=( b r+  c_1)^2,&&B_1=( a y+ {d}_1)^2,\\
    & A_{23}=\ b r^2+2\,{c}_1 r+{ c}_2,&& B_{23}=-(a y^2+2\, {d}_1y+{ d}_2),\label{solForDb}
\end{align}
\end{subequations}
where the constants ${c}_1$, ${d}_1$, ${c}_2$ and ${d}_2$ according to equations (\ref{Killing uslovie}, \ref{sigCons}) are subject to the following condition:
\begin{align}\label{consDop}
     b\,{c}_2+{a}\,{d}_2={c}_1{}{\!^2}+{d}_1{}^{\!2}.
\end{align} 
Only when all these conditions are satisfied does the Killing-Yano tensor become legitimate and finally given by the NP components 
\begin{align} &Y_{ln}=\sqrt{B_1},&&Y_{m\bar m}=-i\sqrt{A_1}.
\end{align}

It remains to check that with these  conditions   the Weyl tensor projections $\Psi_{1,3}$ vanish. Taking into account the conditions (\ref{ab}\, \ref{435}, \ref{sigCons}), $\Psi_{1,3
}$ can be written in the form   
    \begin{align}
\label{PSI13Explicit}
    8\Sigma^3 \Psi_{1,3}&=-\sqrt{A_2B_2}\left\{\Sigma(a   A_{23}'' -b B_{23}'')  -ab( A_{23}'{}^2 +B_{23}'{}^2 )\right\},
\end{align}
Substituting here (\ref{solForDa} \ref{solForDb} \ref{consDop}),
one finds that 
$\Psi_{1,3
}=0$ indeed. 
Also note that in the static case, when $A_5=0=B_3$ and hence $a=0,\, B_{23}=0$, we have $\Psi_{1,3
}=0$ without imposing a condition of type $D$.

The Ricci scalars and the Weyl scalar $\Psi_2$ for generic type $I_B$ are given in the Appendix {\ref{Appendix_B}}.
 \subsection{Conformal Killing tensor}
The conformal Killing tensor (CKT) must satisfy the equations
\begin{align}
\label{Conformal Initial} \nabla_{(\alpha}\mathcal K_{\mu\nu)}=\Omega_{(\alpha}g_{\mu\nu)},\qquad
    6\,\Omega_\alpha=(2\nabla_\sigma \mathcal K_{\alpha}{}^\sigma+\nabla_\alpha \mathcal K),\qquad \mathcal K=g^{\mu\nu}\mathcal{K}_{\mu\nu}.
\end{align}
For  $\Omega_\alpha=0$, these equations in our $I_B$ class of metrics have a solution  with only two NP projections $K_{ln},\,K_{m \bar m}$, so it  is natural to look for CKT of similar structure. Projecting Eq. (\ref{Conformal Initial}) onto the chosen tetrad (\ref{BT}) one gets the following system 
 of four equations
 \begin{align}
 \nonumber
&(D+\rho+\bar\rho)(\mathcal K_{ln}+\mathcal K_{m\bar m})=0,&(\delta+\tau-\bar\pi)(\mathcal K_{ln}+\mathcal K_{m\bar m})=0,\\
&(\Delta-\mu-\bar \mu) (\mathcal K_{ln}+\mathcal K_{m\bar m})=0,
&(\bar \delta+\bar \tau-\pi)(\mathcal K_{ln}+ \mathcal K_{m\bar m})=0.
\end{align}
After substitution of an explicit expression (\ref{AddSpinCoeff1},  \ref{AddSpinCoeff2}) for the spin coefficients, this can be easily integrated. Thus, up to an arbitrary function ${} S(r,y)$, we can write down the NP projections 
\begin{align}
    \mathcal K_{ln}=\Sigma-{} S(r,y),&&\mathcal K_{m\bar m}={} S(r,y).
\end{align}

The conformal factor $\Sigma=\Sigma(r,y)$ here may be non-separable. 
The conformal tensor transforms into an exact one for ${} S=A_1$ and the condition (\ref{Killing uslovie}).

\section{Slice-reducibility and massive particle surfaces}

\subsection{Black hole shadows and characteristic surfaces}
Recent interest to black hole shadows \cite{EventHorizonTelescope:2019dse,EventHorizonTelescope:2022wkp}
gave rise to new theoretical approaches. Strong gravitational lensing, quasi-normal modes of black holes and the formation of black hole shadows are determined by the motion of massless particles near photon surfaces \cite{Claudel:2000yi,Virbhadra:2002ju,Gibbons:2016isj}, on which photons winds before scatter to infinity.  Such surfaces  exist in static metrics, while in rotating  case  similar role is played by surfaces where non-planar spherical orbits are located \cite{Teo:2020sey}. In the first case, the corresponding hypersurfaces in space-time are umbilic \cite{Kobialko:2020vqf} (the tensor of external curvature is proportional to the induced metric), in the second case they are partially umbilic, which means that the latter property is satisfied not for tensors as a whole, but by their convolutions with a part of the vectors of the tangent space. 

In a similar way, one can consider the characteristic surfaces of massive particles, as well as particles of variable mass, for example, photons in plasma \cite{Bogush:2023ojz,Kobialko:2023qzo}. It was noted \cite{Pappas:2018opz,Glampedakis:2018blj,Konoplya:2021slg} that the existence of hypersurfaces with the above properties correlates with the existence of the Killing tensor. 
Eventually, it was shown \cite{Kobialko:2021aqg,Kobialko:2022ozq} that Killing tensors, which reduce to trivial (products of Killing vectors) on hypersurfaces that can be used to stratify the entire spacetime, ensure that these hypersurfaces contain generalized photon surfaces, including those associated with the motion of massive particles.

It is remarkable that  the BF Killing tensor (\ref{Killing}) possesses the slice-reducibility property. As a result, the BF ansatz can be used to give unified description of particle surfaces and shadows of supergravity black holes \cite{Kobialko:2021aqg,Kobialko:2022ozq,Kobialko:2023qzo,Bogush:2023ojz}.
\subsection{Slice-reducibility} 
To see that Benenti Killing tensor (\ref{Killing}) is slice-reducible \cite{Kobialko:2022ozq}, it is enough write it in the form 
\begin{equation}\label{eq:K_r}
   K^{\mu\nu}=
   -A_1 g^{\mu\nu}
   -A_2\delta^\mu_r \delta^\nu_r
   +\tilde K^{\mu\nu}_r,
\end{equation}
where
\begin{equation}\label{KKV1}
\tilde K^{\mu\nu}_r=A_3\delta^\mu_t \delta^\nu_t +2A_4\delta^{(\mu}_t \delta^{\nu)}_\varphi+A_5\delta^\mu_\varphi \delta^\nu_\varphi.  
\end{equation}
The first term in (\ref{eq:K_r})  is trivial Killing tensor on ${\cal S}_r$, since $A_1=\rm const$ there. The second term is orthogonal to ${\cal S}_r$ and thus irrelevant, while the third term $\tilde K^{\mu\nu}_r$ is a reducible Killing tensor on this hypersurface, being presented as linear combination of the tensor products of Killing vectors projected onto it.

Similarly the BF Killing tensor can be presented in terms of ${\cal S}_y$ foliation:
\begin{equation}
\label{eq:K_y}
   K^{\mu\nu}=
   B_1 g^{\mu\nu}
   +B_2\delta^\mu_y \delta^\nu_y
   +\tilde K^{\mu\nu}_y.
\end{equation}
with the slice projection
\begin{equation}\label{KKV2}
\tilde K^{\mu\nu}_y=B_3\delta^\mu_t \delta^\nu_t + 2B_4\delta^{(\mu}_t \delta^{\nu)}_\varphi+B_5\delta^\mu_\varphi \delta^\nu_\varphi.
\end{equation}
If we omit terms with the metric tensor in Eqs. (\ref{eq:K_r}) and (\ref{eq:K_y}), we get simple expressions for conformal Killing tensors, which depend only on one coordinate. \

Thus, the Benenti ansatz ensures the Killing tensor is slice-reducible with respect to both foliations of spacetime. In particular, automorphism (\ref{eq:automorphism}) naturally arises as a symmetry of slices discussed in \cite{Kobialko:2022ozq}. 
\subsection{Photon and massive particle surfaces} 
The black hole horizon $r_h$ is the largest root of the equation
\begin{equation}
  A_2(r_h)=0, \qquad A_2>0 \;\;\mbox{for} \;\;\;r>r_h.   
\end{equation}
In spacetime this is a null hypersurface.
Consider the timelike three-dimensional hypersurface ${\cal S}_r$ for $r={\rm const}>r_h$ with the unit outward normal spacelike covector $n_\mu=- \sqrt{\Sigma/A_2}\,\delta^r_\mu$. The induced metric and the extrincsic curvature of $S_r$ in the bulk coordinates read   
\begin{equation}
h_{\mu\nu}=g_{\mu\nu}+n_\mu n_\nu, \quad  \chi_{\mu\nu}=h^\alpha_\mu h^\beta_\nu \,\nabla_\alpha n_\beta=h^\alpha_\mu h^\beta_\nu \,\left(n_{\beta,\alpha}-\Gamma^\lambda_{\alpha\beta}n_\lambda\right).
\end{equation}
From here one finds:
\begin{equation}
 \chi_{\mu\nu}dx^\mu dx^\nu=\frac{n_r}2 g^{rr} \left( g_{ab,r} dx^a dx^b +g_{yy,r}dy^2\right)   
\end{equation}
The hypersurface ${\cal S}_r$ in the static spacetime contains a photon surface, like $r=3M$ in Schwarzschild metric, which is the loci of confined photon orbits. Its radius  is determined by the umbilicity condition
\begin{equation}  \chi_{\mu\nu}=\frac{\chi^\alpha_\alpha}3 h_{\mu\nu},\qquad \mu,\nu=t,\varphi, y
\end{equation}
which reduces to 
\begin{equation}
    \left(\ln g_{tt}\right)_{,r}=\left(\ln |g_{\varphi\varphi}|\right)_{,r}=\left(\ln |g_{yy}|\right)_{,r}
\end{equation}
Substituting BF coefficients one finds that the last equality is trivially satisfied
(showing that we are dealing with geometrical sphere), while the first one defines the radius of the photon sphere. It may be written as
\begin{equation}\label{PhotonSphereStatic}
(\ln A_2)'= 2(\ln A_{23})',\quad \mbox{or simply}\quad A_3'=0,
\end{equation}
where prime denotes the derivative with respect to $r$. Let's test this equation for the Reissner-Nordstrom metric, in which case $A_2= r^2-2Mr+Q^2,\; A_{23}=r^2$. From Eq. (\ref{PhotonSphereStatic}) one obtains
\begin{equation}
    r^2-3Mr+2Q^2=0\quad \Rightarrow \quad r=\frac32\left( M\pm\sqrt{M^2-8Q^2/9}\right),
\end{equation}
which is indeed a correct expression for the photon spheres in Reissber-Nordstrom spacetime.
One can also look for timelike hypersurfaces such that a particle of mass $\mu$ moving along an initially tangent worldline to ${\cal S}_r$ with tangential vector $p^\mu$ (normalized as $p^\mu p_\nu=\mu^2$) remains there forever.
Such a surface was named
{\em massive particle surface}  (MPS).
The surface radius $r$ will depend on the particle energy and angular momentum integrals $E,\,L$, and the family of MPS  will foliate certain four-dimensional subspace in space-time. The (partial) umbilicity condition in this case must be satisfied only in those directions of the tangent space of ${\cal S}_r$ that are orthogonal to the   Killing vectors.

In the general stationary case, the induced metric and the second quadratic form can be written as
\begin{align}
&h_{\mu\nu}dx^\mu dx^\nu=\frac{A_2}{\Sigma}\left(b dt-B_{23}d\varphi\right)^2-\frac{B_2}{\Sigma} \left(a dt-A_{23}d\varphi\right)^2- \frac{\Sigma}{B_2} dy^2,\quad  \\
    & \chi_{\mu\nu}dx^\mu dx^\nu= \frac12\sqrt{\frac{A_2}{\Sigma}}  \partial_r \,h_{\mu\nu}dx^\mu dx^\nu.
\end{align}

In this case the MPS radius is a solution of the following equation of partial umbilicity \cite{Bogush:2023ojz}:
\begin{equation}
    \frac12 h_{yy}\sqrt{|g^{rr}|}\partial_r p^2 =\chi_{yy}\left(p^2-\mu^2\right),\qquad p^2=g^{ab}p_a p_b,\qquad p_a=(-E,L),
\end{equation}
where $E,\,L$ are constants of motion defined  in (\ref{SEL}).
Substition of BF coefficient functions  leads to 
\begin{equation}\label{MPS}
 \left(\sqrt{A_3}E-\sqrt{A_5}L\right)'   \left(\sqrt{A_3}E-\sqrt{A_5}L\right)=\frac12  \mu^2 A_{1}'  
\end{equation}
where primes denotes derivatives over $r$. This equation  is nothing but $U_r'=0$ for the radial potential defined in (\ref{Ury}). As was noted in Sec. 3, for spherical orbits the Carter constant is determined by two other integrals of motion. This explains why the equation for MPS radius defines it as function of two parameters only. 

Detailed description of spherical orbits of massive particles in Kerr spacetime was given by Teo, \cite{Teo:2020sey}, where the suitable domains of  ${\cal C}$ are determined. More general setting, including the case of charged particles and presence of the electromagnetic field can be found in \cite{Bogush:2023ojz}. In the massless case, spherical orbits are determined by the ratio $L/E$, called the impact parameter, due to the scale invariance of the geodesic equations. 

Our Eq. (\ref{MPS}) is  a general equation defining the MPS radius in terms of the motion integrals $E,\,L$ for the entire class of $I_B$-type metrics. The regions of these parameters  is restricted by consistency conditions, this is discussed in detail in  \cite{Teo:2020sey,Bogush:2023ojz}. The particular cases are described in the next section.
\section{ Supergravity black holes}  
Black hole solutions in ${\cal N}=2,4,8$ supergravity models have been constructed using the Harrison transformations applied to the Kerr metric. The most general results were obtained in this way by Chow and Compere \cite{Chow:2014cca}. Here we show that the known solutions indeed belong to our doubly restricted (\ref{435},\ref{sigCons}) BF class $I_B$. This class ensures the separability of not only the Hamilton-Jacobi and Klein-Gordon equations, but also the Einstein equations \cite{Carter:1968ks}. All metrics also satisfy the gauge condition (\ref{ab}) in Boyer-Lindquist coordinates with the gauge constant $a$ equal to the rotation parameter. Therefore, in this section we will use the $a$ symbol in both senses.

\subsection{Kerr-Newman  }
Kerr-Newman metric may be viewed as type $D$ solution of ${\cal N}=2$ pure supergravity which does not involve scalar fields. In Boyer-Lindquist  coordinates, the  interval reads:
\begin{align*}
    &ds^2=\frc{\Delta_r}{\Sigma}\,(dt-a\sin^2\theta\, d\varphi)^2-\frc{1}{\Sigma}\,\sin^2\theta\,(a\, dt-(r^2+a^2)d\varphi)^2-\frc{\Sigma}{\Delta_r}\,dr^2-\Sigma\,d\theta^2,\\
 &\qquad\qquad\qquad\Delta_r=r^2-2Mr+a^2+Q^2,\quad\Sigma=r^2+a^2\cos\theta,
\end{align*}

The corresponding BF metric  coefficients  $A_k(r)$ and $B_k(y)$ read: 
    \begin{align}
\nonumber
   & A_1=r^2,&&B_1=a^2y^2\\
   \nonumber
   & A_3=\frc{(r^2+a^2)^2}{\Delta_r},&&B_3={a^2({1-y^2})},\\
   &A_5=\frc{a^2}{A_2}=\frc{a^2}{\Delta_r},\,  &&B_5=\frc{1}{B_2}=\frc{1}{(1-y^2)}, 
\end{align}
where $y=\cos\theta$. It is easy to verify that conditions (\ref{435},\ref{sigCons}) are fulfilled:
\begin{equation}
     A_4^2=A_3 A_5,\;\; B_4^2=B_3B_5, \;\;  b=1, \;\; A_1+B_1=\sqrt{A_2 B_2}\left(\sqrt{A_3 B_5}-\sqrt{A_5 B_3}\right).
\end{equation}
The BF formulas for type $D$ (\ref{solForDa},\ref{solForDb},\ref{consDop}) also hold. The non vanishing  Weyl and Ricci scalars for this metric are
\begin{align}
    \Psi_2=\frac{Q^2-M(r-i\,ay)}{(r-i\,ay)^3(r+i\,ay)},&&\Phi_{11}=\frac{Q^2}{2(r^2+a^2)^2}.
\end{align} 
The MPS equation is a 5-th order polynomial
\begin{align}
    \left(E(r^2+a^2)-a L\right)\left [E r\,(\Delta_r-M r+Q^2)+a L(r-m)+a^2 E M\right ]-\mu^2 r\Delta_r{}^2=0,
\end{align}
which can be subjected to numerical analysis.
\subsection{ Gal'tsov-Kechkin solution}
The seven–parametric family of  rotating dilaton–axion–NUT dyons in truncated ${\cal N} =4$ supergravity  obtained  in \cite{Galtsov:1994pd} reads:
\begin{align}\label{GK}
&ds^2=\frac{\Delta_r-a^2 \sin^2\theta}{\Sigma}(dt-w d\varphi)^2-  \Sigma\left(\frac{dr^2}{\Delta_r}+d\theta^2+\frac{\Delta_r\sin^2\theta}{\Delta_r-a^2\sin^2\theta}d\varphi^2\right),\\
\nonumber
   &\Delta_r=(r-r_-)(r-2M)+a^2-(N-N_-)^2,\\
   \nonumber
   &\Sigma=r(r-r_-)+(a\cos \theta+N)^2-N_-^2,\\ \nonumber
   &w=\frac{2}{a^2\sin^2\theta-\Delta_r}\left[N\Delta_r\cos\theta+a\sin^2
   \theta(M(r-r_-)+N(N-N_-))\right],\\ \nonumber
   & r_-=\frac{M|Q-iP|^2}{|M+iN|^2},\qquad\qquad N_-=\frac{N|Q-iP|^2}{2|M+iN|^2}.
\end{align}

The physical parameters are mass $M$, NUT-charge $N$, electric and magnetic charges $Q,\,P$, a rotation parameter $a$, the asymptotic values of dilaton and axion a set zero  (for 
$P=0=N$ this metric transforms to the Sen metric \cite{Sen:1992ua} via the coordinate shift $r\rightarrow r+r_-$).
In the BF form (\ref{MetrUpBeneti}) with $\cos\theta=y$ we have:  
     \begin{align}
\nonumber
   & A_1=r(r-r_-),&&B_1=(a\,y+N)^2-N_-^2,\\
   \nonumber
   & A_3=\frc{\left(r(r-r_-)+a^2+N^2-N_-^2\right)^2}{\Delta_r},&&B_3=\frc{\left[a({1-y^2})-2Ny\right]^2}{1-y^2},\\
   &A_5=\frc{a^2}{A_2}=\frc{a^2}{\Delta_r},\,  &&B_5=\frc{1}{B_2}=\frc{1}{1-y^2},\\
   & A_4^2=A_3 A_5,\;\; B_4^2=B_3B_5, \;\;{} a=a,\;\; {} b=1, &&A_1+B_1=\sqrt{A_2 B_2}\left(\sqrt{A_3 B_5}-\sqrt{A_5 B_3}\right), \nonumber
\end{align}
the last relation being the separability condition (\ref{sigCons}) for the Klein-Gordon equation.  
The metric is a non-vacuum Petrov type $I_B$  solution, with the following   set of non-zero Weyl  and Ricci scalars:
\begin{align}
\nonumber
&\Psi_1=\Psi_3=\frac{ a(4N_-^2+ r_- ^2) \sin \theta  \sqrt{\Delta_r }}{8 \Sigma^3},&\\ \nonumber
12\Sigma^3\Psi_2=&-12 (M-i N) (r+i (a \cos \theta +N))^3+6N N_- \big(2(r+i(N+a \cos \theta))-r_-\big)^2+\\
 \nonumber
 &+r_- ^3 (8 M-r)+8 N N_-^3-4 N_-^4-
6 a r_-  \cos \theta  (5 M-3 i N) (a \cos \theta +2 N-2 i r)+\\
\nonumber
&+
 2 r_-  \left(M \left(15 (r+iN)^2+7 N_-^2\right)+18 N^2 r+9 i N^3+3 i N \left(N_-^2-3 r^2\right)-2 N_-^2 r\right)+
 \\
 \nonumber
 &+4 N_-^2 \left(2 a^2\!-\!a \cos \theta  ( 3 i (M+i N)+a \cos \theta)-5 M r+2 N^2-3 i N(M+ r)\right)+\\
 \nonumber
 &+4N_-^2r^2+r_- ^2 \left(2 a^2-24 i M N-7 N^2+2 N N_--N_-^2\right)-\\ \nonumber
 &-r_- ^2 \left(a \cos \theta  (6 i (4 M-iN)+a \cos \theta )+26 M r-6 i N r-r^2\right),\end{align}\begin{align}
\nonumber 16\Sigma^3\Phi_{11}=&a^2 \cos ^2(\theta ) \left(8 M r_- +16 N N_--4 N_-^2-r_- ^2\right)+16 M N_-^2 r_--24 M N_-^2 r +\\
    \nonumber
    &+8 a N \cos \theta  \left(2 M r_- 
+4 N N_--4 N_-^2-r_- ^2\right)+N^2 \left(8 M r_- -28 N_-^2-7 r_- ^2\right)+\\
\nonumber
&+8 M r_-  r^2-14 M r_- ^2 r+16 N^3 N_-+2 N N_- \left(4 N_-^2+
3 r_- ^2+8 r^2-8 r_-  r\right)+\\ \nonumber
&+N_-^2 r_- ^2-4 N_-^2 r^2+
4 N_-^2 r_-  r+4 N_-^4-r_- ^2 r^2+r_- ^3 r+6 M r_- ^3,\\
\label{PhiDeviators}
   &\Lambda=\frac{R}{24}=-\frac{\left(4 N_-^2+r_- ^2\right)(a^2\sin^2\theta+ \Delta_r)}{48 \Sigma^3}.&
\end{align}
Matter fields ensure that the Carter and d'Alembert operators commute, since the NP-projectors of the Ricci tensor \ref{deviators} are zero.
The MPS equation reads:
\begin{align}
\nonumber
    &E\left(r(r-r_-)+N^2-N_-^2
\right) \Big[(2r-r_-)(r^2-2rr_-+a^2+N^2-2NN_-+3N_-^2)-\\
\nonumber
-&2M(3r^2-rr_-+2r_-^2+N^2-N_-^2)
\Big]-4a^2 E L M+a^4 E^2(2r+2M-r_-)+\\
\nonumber
&+a^2\Big[
2E^2(2(2r-r_-)(N_-^2-NN_-)-2M((r-r_-)^2)-N^2+N_-^2)-L^2(2r-r_--2M)\Big]+\\
\nonumber
&+4 a E L \Big[
M(r-r_-)-N^2+N_-^2+N_-(N-N_-)(2r-r_-)
\Big]-\mu^2\Delta^2(2r-r_-)=0.
    \end{align}

\subsection{SWIP solutions}

Another family of supersymmetric extremal stationary solutions of
${\cal N} = 4,\,D = 4$ supergravity containing a set of electric and magnetic charges satisfying Bogomol'nyi (BPS) bound was obtained in \cite{Bergshoeff:1996gg}. The metric has the same form (\ref{GK}) with
\begin{align}
\nonumber
w & = \frac{2}{\Delta_r
-a^{2}\sin^{2}\theta}
\left(N \Delta_r \cos\theta +a\sin^{2}\theta
\left[m (r -(m + |\Upsilon|) ) +{\textstyle\frac{1}{2}}(|{\cal
M}|^{2} -|\Upsilon|^{2}) \right] \right)\, ,
\\
\nonumber
\Delta_r&  =  r[r -2(m+|\Upsilon|)] +a^{2}
+(m+|\Upsilon|)^{2}\, ,
\\
\Sigma & =  r (r -2 |\Upsilon|)
+(a\cos\theta +N)^{2}\, .
\end{align}
The solution depends on complex mass $\mathcal M=m+iN$, axion-dilaton charge \mbox{$\Upsilon=-\frac{2}{\mathcal M}\sum_n\left[\bar\Gamma_{(n)}\right]^2$}, and electromagnetic charges \mbox{$\Gamma_{(n)}=\frac{1}{2}(Q_{(n)}+iP_{(n)})$}
,  $n=1,2$. The BPS identity   reads:
\begin{align}
    |\mathcal M|^2+|\Upsilon|^2-4\sum\limits_n|\Gamma_{(n)}|^2=0.
\end{align}

 In this case
     \begin{align}
\nonumber
   & A_1=r(r-2|\Upsilon|),&&B_1=N+a y,\\
   \nonumber
   & A_3=\frc{(r(r-2|\Upsilon|)+N^2+a^2)^2}{\Delta_r},&&B_3=\frc{(a(1-y^2)-2Ny)^2}{1-y^2},\\
   &A_5=\frc{a^2}{A_2}=\frc{a^2}{\Delta_r},&&B_5=\frc{1}{B_2}=\frc{1}{(1-y^2)}.
\end{align}
 
The non-zero NP quantities are:
\begin{align}
\nonumber
&\Psi_1=\Psi_3=\frac{a\sin\theta|\Upsilon|^2\sqrt{\Delta_r}}{2\Sigma^3},\\
\nonumber
    &3\Sigma^3\Psi_2=3(m-iN)(m-r-ia\cos\theta)(r+i(N+a\cos\theta))^2+\\
    \nonumber
    &\qquad\quad+
    3(m-iN)(3(r+ia\cos\theta)-2m+iN)(r+i(N+a\cos\theta))|\Upsilon|-\\
    \nonumber
    &\qquad\quad-2|\Upsilon|^2(3N^2-2(m^2+r^2)+7mr+6iN(m-r)+
    a^2\cos{2\theta})-\\
    \nonumber
    &\qquad\quad-2|\Upsilon|^3
    (4(r-m)+3i(N+a\cos\theta))+3ia|\Upsilon|^2\cos\theta(2(m-iN)-r)+4|\Upsilon|^4,\\
\nonumber
    &4 \Sigma^3\Phi_{11}=3|\Upsilon|^2(a^2\cos^2\theta+(m-r+|\Upsilon|)^2)+\\
    \nonumber
    &\qquad\;\quad+2\Sigma(m^2-a\cos\theta(2N+a\cos\theta)-(r-|\Upsilon|)^2+\Sigma),\\
    \nonumber
    &\Phi_{00}=\Phi_{22}=\frac{|\Upsilon|^2 \Delta_r}{2 \Sigma^3},\qquad\Phi_{02}=-\frac{a^2|\Upsilon|^2\sin^2\theta}{2\Sigma^3}, \\
    &\Lambda=-\frac{|\Upsilon|(a^2(1+\sin^2\theta)+(m-r+|\Upsilon|)^2)}{12\Sigma^3}.
\end{align}
Since $\Phi_{01}=\Phi_{12}=0$,
the separability condition for the Klein-Gordon equation aldo holds. 

The MPS equation reads:
\begin{align}
\nonumber
    &E(r(r-2|\Upsilon|)+N^2+a^2)\Big[\big\{E
    (\Delta_r+|\Upsilon|^2-M^2-N^2)+aL\big\}(r-M-|\Upsilon|)+2a^2EM
    \Big]-\\
    -&\mu^2\Delta_r^2(r-|\Upsilon|)=0.
\end{align}
    
\subsection{STU black holes}
The STU solution is a general asymptotically flat, stationary black hole in supergravity ${\cal N}=8$ \cite{Chow:2014cca}, parameterized by mass $M$, rotation parameter $a$, and four electric charges $s_I$, $I=1,..,4$. The metric has the same form (\ref{GK}) with
\begin{align*}
   &\Delta_r=r^2-2Mr+a^2,\\
   &w=-\frac{2Ma\omega\sin^2\theta}{\Delta_r-a^2\sin^2\theta},\qquad\omega=((\Pi_c-\Pi_s)r+2M\Pi_s),\\
   &\Sigma^2=\prod\limits_{I=0}^4(r+2M s_I^2)+a^4\cos^4\theta+\\
   &\quad\;
   +2a^2\cos^2\theta\left(r^2+Mr\sum\limits_{I=0}^3s_I^2+4M^2(\Pi_c-\Pi_s)\Pi_s-2M^2\sum\limits_{I<J<K}^3s_I^2s_J^2s_K^2\right),
\end{align*}
and the products of charges are defined as follows
\begin{align*}
    \Pi_s=\prod\limits_{i=1}^4s_I=\prod\limits_{I=1}^4\sinh \delta_I,\qquad\Pi_c=\prod\limits_{I=1}^4\sqrt{1+s_I^2}=\prod\limits_{I=1}^4\cosh \delta_I,&&s_I^2=\sinh^2\delta_I.
\end{align*}

In the case of pairwise equality of charges $s_1=s_3=\mathcal S_1$, $s_2=s_4=\mathcal S_2$ we are dealing with the so-called two-charge solution. If $\mathcal S_2=0$, then the metric reduces to the Kerr-Sen solution~\cite{Sen:1992ua}. For a more general classification, see \cite{Cvetic:2017zde}.
It is possible to transform the metric to the BF form (\ref{MetrUpBeneti}) only in the case of a two-charge solution, for which
\begin{align}
    \Sigma_{2ch}=r^2+a^2\cos^2\theta+2Mr(\mathcal S_1^2+\mathcal S_2^2)+4M^2\mathcal S_1^2\mathcal S_2^2=r^2-2Mr+a^2\cos^2\theta+2Mw_{2ch}
\end{align}

Then for the functions $A_i(r)$, $B_i(x)$ we have:
     \begin{align}
\nonumber
   & A_1=r^2-2Mr+2Mw_{2ch},&&B_1=a^2y^2,\\
   \nonumber
   & A_3=\frc{(r^2-2Mr+a^2+2Mw_{2ch})^2}{\Delta_r},&&B_3=a^2(1-y^2),\\
   &A_5=\frc{a^2}{A_2}=\frc{a^2}{\Delta_r},&&B_5=\frc{1}{B_2}=\frc{1}{(1-y^2)}.
\end{align}
The Weyl and Ricci scalars are
\begin{align}
\nonumber
 &\Psi_1=\Psi_3=\frac{ a M^2\sin\theta(S_1^2-S_2^2)^2\sqrt{\Delta_r}}{2\Sigma^3_{2ch}},&\\
\nonumber
\frac{3\Sigma^3_{2ch}}{M}\Psi_2&=
 -3(r+ia\cos\theta)^3-3(r+ia\cos\theta)^2(M+r+ia\cos\theta)(\mathcal S_1^2+\mathcal S_2^2)-\\
 \nonumber
 &-2M(r^2+2Mr-a^2+3iar\cos\theta+a^2y^2)(\mathcal S_1^4+\mathcal S_2^4)+\\
 \nonumber
 &+12M^2\mathcal S_1^2\mathcal S_2^2(M-r-ia\cos\theta)(\mathcal S_1^2+\mathcal S_2^2)+\\ \nonumber
 &+
 4M\mathcal S_1^2\mathcal S_2^2(5a^2\cos^2\theta-a^2+4Mr-5r^2+3iaM\cos\theta-9iar\cos\theta),\\
\nonumber &\Phi_{00}=\Phi_{22}=\frac{M^2\Delta_r(\mathcal{S}_1^2-\mathcal{S}_2^2)^2}{2\Sigma^3_{2ch}},\qquad\Phi_{02}=\overline{\Phi_{20}}=-
    \frac{a^2 M^2\sin^2\theta(\mathcal{S}_1^2-\mathcal{S}_2^2)^2}{2\Sigma^3_{2ch}},\\
\nonumber  \frac{4\Sigma^3_{2ch}}{M^2}\Phi_{11}&=4(r^2+a^2\cos^2\theta)(\mathcal S_1^2+\mathcal S_2^2)+(2M r+3r^2+3a^2\cos^2\theta)(\mathcal S_1^4+\mathcal S_2^4)+\\ \nonumber
    &+16M(r+M)\mathcal S_1^2\mathcal S_2^2(\mathcal S_1^2+\mathcal S_2^2)+32M^2\mathcal S_1^4\mathcal S_2^4+2\mathcal S_1^2\mathcal S_2^2(r^2+14Mr+a^2\cos^2\theta),\\
    &\Lambda=\frac{1}{24R}=-\frac{M^2 (\mathcal{S}_1^2-\mathcal{S}_2^2)^2 \left(\Delta_r+a^2 \sin ^2\theta \right)}{12 \Sigma^3_{2ch}}.&
\end{align}
Similar to the above cases of ${\cal N}=4$, the quantum separability condition (\ref{KommutatorKilling}) is also satisfied for the two-charge STU solution.

The MPS equation reads
\begin{align}
\nonumber
    &E^2\Big[\Delta^2(r-M+2M\Sigma_{s^2})-4M^2 r(Mr-a^2)\Sigma_{s^2}^2\!-\!16M^4(r-M)\Pi_{s^2}^2\!-\!8M^3(r^2-a^2)\Pi_{s^2}\Sigma_{s^2}\Big]+\\
    +&2aELM\Big[4M(r-M)\Pi_{s^2}-(r^2-a^2)\Sigma_{s^2}\Big]\!-\!a^2L(r-M)-\mu^2\Delta^2(r-M+M\Sigma_{s^2})=0,
\end{align}
where
$   \Sigma_{s^2}=\mathcal{S}_1^2+\mathcal{S}_2^2,\;\Pi_{s^2}=\mathcal{S}_1^2\mathcal{S}_2^2.
$
 
From the above analysis of the various supergravity solutions for black holes, their similarity in the form of BF is obvious. For all of them, the coefficient functions $A_1, B_1, A_{23}, B_{23}$ are just simple quadratic polynomials, which make us to believe that these solutions can be obtained by directly integrating the general supergravity equations.
\section{Conclusions}
We have shown that the suitably refined Benenti-Francaviglia ansatz for stationary axisymmetric spacetimes, admitting a non-trivial Killing tensor of rank two, defines a class of non-algebraically special metrics that admits two null geodesic congruences without shear. If we additionally impose the separability condition of the Klein-Gordon equation, the resulting ansatz coincides with Carter's parametrization, which led him to perform a direct integration of Einstein's vacuum and electrovacuum equations, reproducing, in particular, the Kerr-Newman metric and its extensions of type D. Our class of algebraically non-special metrics opens the way to a generalization of this approach to the general supergravity bosonic action  (\ref{eq:general_action}) containing a set of vector and scalar fields. It was   demonstrated explicitly that all known ${\cal N}=2,\,4,\,8$ supergravitational black holes, with the exception of the so-called four-charge solution, belong to our class indeed.  

An algebraically special subclass of our metrics contains only the type
$D$. For this case, we have obtained an explicit parametrization of the BF coefficient functions integrating the Killing-Yano equations.

Our class of metrics leads to a general description of the confining surfaces of photons and massive particles that have recently been introduced in the theory of black hole shadows. The Killing tensor given by the BF formula has a crucial property of slice-reducibility  \cite{Kobialko:2022ozq} with respect to spacetime bundles by three-dimensional timelike hypersurfaces with constant $r$ or $y$. 
This opens a way to universal description of supergravity black hole shadows. Here we have given a general equation for the confining surfaces of massive particles and its particular form for all examples of supergravity solutions.

To summarize, the restricted Benenti-Francavilla metric gives a unified description of type I supergravity black holes, revealing many of their common properties. This was established by combining the Newman-Penrose analysis of the BF class of metrics with a specific form of known solutions. We hope that it will be possible to directly solve the supergravity equations of motion in the obtained parameterization, thus extending Carter's program to solve the Einstein equations for metrics admitting second-rank Killing tensors.

\section*{Acknowledgements} 
The authors thanks Kirill Kobialko, Igor Bogush and G\'erard Cl\'ement for useful suggestions and discussions. This work was supported by Russian Science Foundation under Contract No. 23-22-00424.

\appendix
\section{Spin coefficients}
Here we give NP spin coefficients calculated for the tetrad (\ref{BT}) without constraint (\ref{sigCons}) on the conformal factor $\Sigma$:
\begin{align}
\label{AddSpinCoeff1}
  &\mu=\rho=-
\sqrt{A_2}\,\frac{\partial}{\partial_r}\frac{1}{\sqrt{2\Sigma}}-\frac{iB_5 \sqrt{B_2}}{2 \sqrt{2\,\Sigma }} \cfrac{\partial_y\sqrt{B_3/B_5}}{ \sqrt{A_3B_5 }   -\sqrt{A_5B_3 } },&&\\
  \label{AddSpinCoeff2}
  &\tau=\pi=-i\sqrt{B_2}\,\frac{\partial}{\partial_y}\frac{1}{\sqrt{2\Sigma}}+\cfrac{A_5\sqrt{A_2}}{2 \sqrt{2\, \Sigma} } \frac{\partial_r\sqrt{A_3/A_5} }{ \sqrt{A_3B_5 }   -\sqrt{A_5B_3 } },&&\\
  \label{AddSpinCoeff3}
&\epsilon=\gamma=\frac{1}{2}\,\rho+\frac{\sqrt{A_2}}{2\sqrt{2}\Sigma^{3/2}}\left(\Sigma\,\partial_r\ln\left[\sqrt{A_3 B_5}-\sqrt{A_5 B_3}\right]-\partial_r\Sigma
  \right),&&\\[0em]
  \label{AddSpinCoeff4}
  &\alpha=\beta=\frac{1}{2}\,\tau+i\frac{\sqrt{B_2}}{2\sqrt{2}\Sigma^{3/2}}\left(\Sigma\,\partial_y\ln\left[\sqrt{A_3 B_5}-\sqrt{A_5 B_3}\right]-\partial_y\Sigma
  \right).&&
\end{align}

\section{Ricci and Weyl scalars}
\label{Appendix_B}
With account for the second constraint the NP projections of te Ricci and Weyl scalars are
\begin{align}
 \label{Phi00}
 &\Phi_{00}=\Phi_{22}=\frac{bA_2}{8\Sigma^3}\left(b A_{23}'{}^2+ bB_{23}'{}^2-2\Sigma\,A_{23}''\right),\\ 
 \label{Phi01}
&\Phi_{01}=\overline{\Phi_{10}}=\Phi_{12}=\overline{\Phi_{21}}=
 -\frac{\sqrt{A_2B_2}}{8\Sigma^3}\left( a\,A_{23}''- b\,B_{23}''\right),\\ &\Phi_{02}=\overline{\Phi_{20}}=-\frac{aB_2}{8\Sigma^3}\left(
  a A_{23}'{}^2
+a B_{23}'{}^2+2\Sigma\,B_{23}''\right),
\end{align}
\begin{align}
\nonumber
64\Sigma^3\Phi_{11}&=\Sigma^2\Big\{
A_2\!\left[\big(\!\ln\! A_5\big){}_{\!,r}{\!\!}^2-4\big(\!\ln \!A_5\big){}_{\!,rr}\right]-B_2\!\left[\big(\!\ln\! B_5\big){}_{\!,r}{\!\!}^2-4\big(\!\ln \!B_5\big){}_{\!,rr}\right]
\Big\}+\\
\nonumber
&+3\Big\{
\!4 b^2 \!A_2B_{23}'{}^2-\!4 a^2 \!B_2A_{23}'{}^2+\!A_2\big[\Sigma(\ln \!A_5)_{\!,r}\!+\!2b A_{23}'\big]^2\!-\!B_2\big[\Sigma(\ln \!B_5)_{\!,y}\!-\!2a B_{23}'\big]^2\!\Big\}+\\
\nonumber
&+4\Sigma\Big\{
A_2((\ln A_5)_{\!,r})\big[\Sigma(\ln \!A_5)_{\!,r}\!+\!2bA_{23}'\big]\!-\!A_2\big[\Sigma(\ln \!A_5)_{\!,r}\!+\!2bA_{23}'\big]{}_{\!,r}
\Big\}-\\
&-4\Sigma\Big\{
B_2((\ln B_5)_{\!,y})\big[\Sigma(\ln \!B_5)_{\!,y}\!-\!2a B_{23}'\big]\!-\!B_2\big[\Sigma(\ln \!B_5)_{\!,y}\!-\!2a B_{23}'\big]{}_{\!,y}
\Big\},
\end{align}
\begin{align}
\nonumber
    192\, \Sigma^3\Lambda=8\,\Sigma^3 R&=3\Sigma^2\left\{
    A_2\big[3(\ln A_5){}_{\!,r}{}^{\!\!2}-4(\ln A_5){}_{\!,rr}\big]+B_2\big[3(\ln B_5){}_{\!,y}{}^{\!\!2}-4(\ln B_5){}_{\!,yy}\big]
    \right\}+\\
    \nonumber
    &+4A_2\big[\Sigma(\ln A_5){}_{\!,r}+2b A_{23}'\big]{}_{\!,r}+4B_2\big[\Sigma(\ln B_5){}_{\!,y}-2a B_{23}'\big]{}_{\!,y}-4a^2 B_2\, A_{23}'{}^2-\\
    &-A_2\big[\Sigma(\ln A_5){}_{\!,r}+2b A_{23}'\big]^2-B_2\big[\Sigma(\ln B_5){}_{\!,y}-2a B_{23}'\big]^2-4 b^2 A_2 \,B_{23}'{}^2.
\end{align}
\begin{align}
\label{PSI2Explicit}
\nonumber
     -12\Sigma^3\Psi_2&={\Sigma}\Big\{A_2\Big[\Sigma\big(\ln\!A_5\big){}_{\!,r}\!+\!2b A_{23}'\Big]{}_{\!,r}\!+\!B_2\Big[\Sigma\big(\ln{\!B_5}\big){}_{\!,y}-2aB_{23}'\Big]{}_{\!,y}\Big\}\!+2_{\,}a^2 B_2\,A_{23}'{}^2+\\
    \nonumber
    &-\!A_2\Big[\Sigma\big(\ln\!A_5\big){}_{\!,r}\!+\!2b A_{23}'\Big]^2\!-\!B_2\Big[\Sigma\big(\ln{\!B_5}\big){}_{\!,y}-2a B_{23}'\Big]^2\!+2_{\,}b^2A_2\,B_{23}'{}^2+\\    
    &+3i\left\{a\,B_2\,A_{23}'\Big[\Sigma\big(\ln{\!B_5}\big){}_{\!,y}-2aB_{23}'\Big]+b\,A_2\,B_{23}'\Big[\Sigma\big(\ln{\!A_5}\big){}_{\!,r}+2b A_{23}'\Big]\right\}.
\end{align}

\end{document}